\documentstyle[12pt,epsfig]{article}
\newcommand{\beq}{\begin{equation}}
\newcommand{\eeq}{\end{equation}}
\newcommand{\beqa}{\begin{eqnarray}}
\newcommand{\eeqa}{\end{eqnarray}}
\newcommand{\beqan}{\begin{eqnarray*}}
\newcommand{\eeqan}{\end{eqnarray*}}
\newcommand{\ba}{\begin{array}}
\newcommand{\ea}{\end{array}}
\newcommand{\no}{\nonumber}

\newcommand{\ol}{\overline}
\newcommand{\ra}{\rightarrow}

\newcommand{\vp}{\varphi}

\newcommand{\wt}{\widetilde}
\newcommand{\wh}{\widehat}

\newcommand{\cL}{{\cal L}}

\newcommand{\cO}{{\cal O}}

\newcommand{\st}{\stackrel}
\newcommand{\dfrac}{\displaystyle \frac}

\newcommand{\co}{\; \; ,}
\newcommand{\nn}{\nonumber \\}
\newcommand{\scs}{\co \;}

\newcommand{\per}{ \; .}
\newcommand{\bea}{\begin{eqnarray}}
\newcommand{\eea}{\end{eqnarray}}
\newcommand{\del}{\partial}
\newcommand{\fsl}{\not\!}

\normalsize
\begin{document}
\begin{titlepage}
\begin{flushright}
CERN-TH/98-197\\
UWThPh-1998-03\\
BUTP-98/13\\
\end{flushright}
\vspace{2.5cm}
\begin{center}
{\Large \bf The One-Loop Functional as a Berezinian*}\\[40pt]
H. Neufeld$^{1,2}$, J. Gasser$^{3}$ and G. Ecker$^{2}$

\vspace{1cm}
${}^{1)}$ CERN, CH-1211 Geneva 23, Switzerland \\[10pt]

${}^{2)}$ Institut f\"ur Theoretische Physik, Universit\"at
Wien\\ Boltzmanngasse 5, A-1090 Vienna, Austria \\[10pt]

${}^{3)}$ Institut f\"ur Theoretische Physik, Universit\"at
Bern\\ Sidlerstrasse 5, CH-3012 Berne, Switzerland

\vfill
{\bf Abstract} \\
\end{center}
\noindent
We employ notions familiar from supersymmetry for constructing the
one-loop functional of general quantum field theories with bosons and
fermions (spin $\leq$ 1/2). To demonstrate the advantages of
such an approach for calculating one-loop divergences, we analyse a
simple Yukawa theory with two different versions of a super-heat-kernel 
expansion. These methods also simplify the calculation of
one-loop divergences for non-renormalizable meson--baryon Lagrangians
occurring in chiral perturbation theory.

\vfill
\begin{flushleft}
CERN-TH/98-197\\
June 1998\\
\end{flushleft}

\noindent * Work supported in part by Schweizerischer Nationalfonds
and by TMR, EC-Contract No. ERBFMRX-CT980169 (EURODA$\Phi$NE) 

\end{titlepage}
\addtocounter{page}{1}
\paragraph{1.}
The one-loop functional of a general bosonic or fermionic
theory can be expressed in terms of the determinant of a second-order
differential operator. A common procedure in theories with
both bosonic and fermionic degrees of freedom is to first integrate
out the fermions and then treat the resulting bosonic theory.

The purpose of this note is to demonstrate the usefulness of treating
bosons and fermions on the same footing. Although this has been common
practice in supersymmetric theories from the early days of SUSY, we
emphasize the advantages of such an approach also for
non-supersymmetric quantum field theories with arbitrary numbers
of bosonic and fermionic fields. Here, we consider
only fields with spin $\leq$ 1/2. 

The method turns out to be especially useful for calculating 
the divergent parts of one-loop functionals occurring in the
renormalization program of chiral perturbation theory with mesons
and baryons \cite{GSS88,Kr90,JM91,BKKM}. In the standard approach, 
the bosonic loop, the fermionic loop and the mixed loop (boson and
fermion lines in the loop) are treated separately
\cite{Eck94,MM96}. This requires a cumbersome investigation of the
singular behaviour of products of
propagators, because the mixed loop does not have the form of a
determinant, as the purely bosonic or fermionic loops. The
SUSY-inspired
treatment, on the other hand, reduces the problem to simple matrix
manipulations, in complete analogy to the familiar
heat-kernel expansion technique for bosonic or fermionic loops.         

We demonstrate the simplicity of this approach in the case of
a Yukawa theory, where the divergences can
relatively easily be extracted by standard Feynman diagram techniques.
 With  applications to baryon 
chiral perturbation theory in mind \cite{NEUF98},  we will
actually analyse the Yukawa theory in two different ways. In the first
method, we square the fermionic differential operator in the usual
way to arrive at the superspace version of an elliptic
second-order differential operator (in Euclidean space).
In the second method, we set up the heat-kernel expansion directly for the
original supermatrix. Since we are not aware of a discussion of this
technique in the literature, we will be more explicit for the second
method. 

\paragraph{2.} We start from a general Euclidean action
\beq
\label{action}
S[\vp,\psi,\ol \psi] = \int d^dx ~ \cL(\vp,\psi,\ol \psi)
\eeq
for $n_B$ real scalar fields $\vp_i$
and $n_F$ spin 1/2 fields $\psi_{\alpha}$. To construct the generating
functional $Z$ of connected Green functions, we couple these
fields to external
sources $j_i$ $(i = 1,\ldots,n_B)$, $\rho_{\alpha}$, $\ol \rho_{\alpha}$
$(\alpha = 1,\ldots,n_F)$,
\beqa
\label{funcint}
e^{-Z[j,\rho,\ol \rho]} =
\int [d\vp d\psi d\ol \psi] \,
e^{- S[\vp,\psi,\ol \psi] + j^T \vp + \ol \psi \rho + \ol \rho \psi}~,
\eeqa
where we have used the notation
\beq
\label{notation}
j^T \vp + \ol \psi \rho + \ol \rho \psi := \int d^dx ~ ( j_i \vp_i +
\ol \psi_{\alpha} \rho_{\alpha} + \ol \rho_{\alpha}\psi_{\alpha})~. 
\eeq
The normalization of the functional integral is determined by
the condition $Z[0,0,0] = 0$.
We denote by $\vp_{\rm cl}$, $\psi_{\rm cl}$
the solutions of the classical equations of motion
\beq
\frac{\delta S}{\delta \vp_i} = j_i, \quad
\frac{\delta S}{\delta \ol \psi_{\alpha}} = \rho_{\alpha}, \quad
\frac{\delta S}{\delta \psi_{\alpha}} = - \ol \rho_{\alpha}~.
\label{EOM}
\eeq
With fluctuation fields $\xi,\eta$ defined by
\beqa
\label{ff}
\vp_i &=& \vp_{{\rm cl},i} + \; \xi_i \no \\
\psi_{\alpha} &=& \psi_{{\rm cl},{\alpha}} + \eta_{\alpha} ~,
\eeqa
the integrand in (\ref{funcint}) is expanded in terms of
$\xi,\eta,\ol \eta$. 
The resulting loop expansion of the generating functional
\bea
Z = Z_{L = 0} + Z_{L = 1} + \ldots \no
\eea
starts with the classical action in the presence of external
sources:
\beq
Z_{L = 0} = S[\vp_{\rm cl},\psi_{\rm cl},\ol \psi_{\rm cl}]
- j^T \vp_{\rm cl} - \ol \psi_{\rm cl} \rho - \ol \rho \psi_{\rm cl}~,
\eeq
where the classical fields are fixed by the external
sources through (\ref{EOM}).
The one-loop term $Z_{L = 1}$ is given by a Gaussian functional
integral
\beqa
\label{Gauss}
e^{-Z_{L = 1}} =
\int [d\xi d\eta d\ol \eta] ~
e^{- S^{(2)}[\vp_{\rm cl},\psi_{\rm cl}, \ol \psi_{\rm cl};
\xi, \eta,\ol \eta]}~,
\eeqa
where
\beq
S^{(2)}[\vp_{\rm cl},\psi_{\rm cl}, \ol \psi_{\rm cl}; \xi, \eta,\ol \eta] =
\int d^dx ~ \cL^{(2)}(\vp_{\rm cl},\psi_{\rm cl}, \ol \psi_{\rm cl};
\xi, \eta,\ol \eta)
\eeq
is quadratic in the fluctuation variables.
Employing the notation introduced in (\ref{notation}), $S^{(2)}$
takes the general form
\beqa
\label{Lfluc}
S^{(2)} &=& \dfrac{1}{2}\xi^TA\xi + \ol \eta B \eta + \xi^T \ol \Gamma \eta +
\ol \eta \Gamma \xi \nn
& = & \dfrac{1}{2}\left(\xi^T A\xi + \ol \eta B \eta - 
\eta^T B^T\ol\eta^T + \xi^T \ol \Gamma \eta - \eta^T \ol \Gamma^T \xi +
\ol \eta \Gamma \xi -  \xi^T \Gamma^T \ol\eta^T \right) ~,
\eeqa
where $A,B,\Gamma$ are operators in the respective spaces;
$A = A^T$  and $B$ are bosonic differential operators, whereas $\Gamma$
and
$\ol \Gamma$ are fermionic (Grassmann) operators.  
They all depend on the classical solutions $\vp_{\rm cl}$, $\psi_{\rm cl}$.
The second explicitly symmetric form of $S^{(2)}$ in (\ref{Lfluc}) will
be used for the definition of the supermatrix operator $K$ in (\ref{KSUSY}). 

The standard procedure for the evaluation of (\ref{Gauss}) is to
integrate first over the fermion fields
$\eta,\ol \eta$ to yield the bosonic functional integral
\beqa
e^{-Z_{L = 1}} =
 \det B \int [d\xi] ~
e^{- \frac{1}{2}  \xi^T(A - \ol \Gamma B^{-1} \Gamma + 
\Gamma^T B^{-1T} \ol \Gamma^T)\xi}~.\nonumber
\eeqa
In this way, we obtain the familiar result
\beqa
\label{Z1}
Z_{L = 1} &=& \dfrac{1}{2} \left[\ln \det (A - \ol \Gamma B^{-1} \Gamma + 
\Gamma^T B^{-1T} \ol \Gamma^T) -\ln \det A_0\right]
- (\ln \det B - \ln \det B_0) \nn
&=& \dfrac{1}{2} \mbox{ Tr } \ln \dfrac{A}{A_0}
- \mbox{Tr } \ln \dfrac{B}{B_0} + \dfrac{1}{2}
\mbox{ Tr } \ln(1 - A^{-1} \ol \Gamma B^{-1} \Gamma 
+ A^{-1}\Gamma^T B^{-1T} \ol \Gamma^T) \no \\
&=& \dfrac{1}{2} \mbox{ Tr } \ln \dfrac{A}{A_0}
- \mbox{Tr } \ln \dfrac{B}{B_0} -
\sum_{n=1}^\infty \dfrac{1}{2n}\mbox{Tr }\left(A^{-1} \ol \Gamma B^{-1} \Gamma
- A^{-1} \Gamma^T B^{-1T} \ol \Gamma^T \right)^n ~,
\label{eqconv}
\eeqa
where                                                             
\bea
A_0 := A|_{j = \rho = \ol \rho = 0}, \quad
B _0 := B|_{j = \rho = \ol \rho = 0} \no
\eea
denote the free-field limit of $A$ and $B$, respectively.
Recalling that $A^{-1}, B^{-1}$ are the scalar and fermion matrix
propagators in the presence of external sources, the one-loop functional
$Z_{L = 1}$ is seen to be a sum of the bosonic one-loop functional
$\frac{1}{2} \mbox{ Tr }\ln (A/A_0)$, the fermion-loop functional
$- \mbox{Tr } \ln (B/B_0)$ and a mixed one-loop functional where scalar and
fermion propagators alternate. In order to determine the
ultraviolet divergences that occur in the last term in
(\ref{eqconv}),
 the calculational inconveniences mentioned
in paragraph 1 are encountered.

In the remainder of this letter, we
discuss a  procedure that allows us to identify the singular
pieces in $Z_{L = 1}$ more directly. Inspired by supersymmetric quantum field 
theories, we  reorganize
the three parts of $Z_{L = 1}$ into a more compact form, using the notion
of
supermatrices, supertraces, etc. (cf., e.g.,
Refs.~\cite{ANZ,GGRS83,berezin}).

Assembling the bosonic
and fermionic fluctuation fields into a multicomponent field
\beqa
\lambda^T = \left(\xi^T, \eta^T, \ol \eta \right) ,\nonumber
\eeqa
$S^{(2)}$ in (\ref{Lfluc}) can be written as
\beqa
S^{(2)} = \dfrac{1}{2} \lambda^T\; K \; \lambda\scs \nonumber
\eeqa
where $K$ is the supermatrix operator 
\beq
K = \left( \ba{ccc}
A & \ol \Gamma & -\Gamma^T \\
- \ol \Gamma^T & 0 & - B^T \\
\Gamma & B & 0
\ea \right)~.
 \label{KSUSY}
\eeq
For a general supermatrix of the form
\beqa
M = \left( \ba{cc} a & \alpha \\ \beta & b \ea \right)\scs\nonumber
\eeqa
where $a,b$ ($\alpha,\beta$) are bosonic (fermionic) variables, one
defines the
supertrace str~$M$ and the Berezinian sdet~$M$ as
\cite{ANZ,GGRS83,berezin}  
\bea
\mbox{str } M &:=& \mbox{tr } a - \mbox{tr } b\scs\nn
\mbox{sdet } M &:=& \det (a - \alpha b^{-1} \beta)/\det b \no~.
\eea
These definitions give rise to the relation
\beq
\mbox{sdet } M = \exp{(\mbox{str }\ln M)}\scs
\eeq
in analogy to the one for ordinary matrices.

Comparing this with Eq. (\ref{Z1}), we obtain
\beq
Z_{L = 1} = \frac{1}{2} \mbox{ Str } \ln \frac{K}{K_0} =
\frac{1}{2} \mbox{ Str } \ln \frac{K'}{K'_0}
\label{Zsuper}
\eeq
with
 \beq
K' = \left( \ba{ccc}
A & \sqrt{\mu} ~ \ol \Gamma & -  \sqrt{\mu} ~ \Gamma^T \\
\sqrt{\mu} ~  \Gamma & \mu B & 0  \\
\sqrt{\mu} ~ \ol \Gamma^T & 0 & \mu B^T
\ea \right)~.
 \label{KP}
\eeq
With our notation
\bea
\mbox{Str } O = \int d^dx ~ \mbox{str} \langle x|O|x \rangle \no
\eea
we are distinguishing supertraces with and without integration
over Euclidean space. In (\ref{KP}) we have introduced
a mass parameter $\mu$ that guarantees equal dimensions for all entries
in $K'$ ($[K']=[A]=2$). Although this quantity does, of course, not appear
in any final result,  it turns out to be quite helpful for the inspection
of expressions at intermediate stages of calculations.

It is seen that the three
types of one-loop functionals (bosonic, fermionic and mixed) in
Eq. (\ref{Z1}) are not distinguished any more in
the representation
(\ref{Zsuper}), where $Z_{L = 1}$ is expressed in terms of a Berezinian.
Generalizing the heat-kernel expansion to supermatrix operators,
the determination of  the divergent part of $Z_{L = 1}$ reduces to
matrix algebra, in complete analogy to the purely bosonic or fermionic
loop functionals.

\paragraph{3.}
As long as we are only interested in those parts of the one-loop
functional that are at most bilinear in fermion fields, we can reduce
the supermatrix $K'$ to the simpler form
 \beq
K'' = \left( \ba{cc}
A & \sqrt{2\mu}~\ol \Gamma  \\
\sqrt{2\mu} ~ \Gamma & \mu B 
\ea \right) \scs
 \label{KPP}
\eeq
such that the one-loop functional can be written as
\beq
Z_{L=1} = \dfrac{1}{2} \mbox{ Str } \ln \dfrac{K''}{K''_0}
- \dfrac{1}{2} \mbox{ Tr } \ln \dfrac{B}{B_0}
 + \ldots \label{Zss}
\eeq
The terms omitted are at least quartic in the fermion fields.

The representation (\ref{Zss}) can be written in a more compact form
by ``squaring'' the fermionic differential operator $B$.  
We first note that
\beqa
\mbox{Tr } \ln \dfrac{B}{\mu} = \ln \det \dfrac{B}{\mu} = 
\ln \det \gamma_5 \dfrac{B}{\mu} \gamma_5 =
\mbox{Tr } \ln \gamma_5 \dfrac{B}{\mu} \gamma_5 
 = -\mbox{Str }
 \ln
\left( \ba{cc} 1 & 0 \\ 0 & \gamma_5 \dfrac{B}{\mu} \gamma_5 
\ea \right). \nonumber
\eeqa
Suppressing terms of higher than second order in the fermion
fields in the rest of this paper, we get
\renewcommand{\theequation}{\arabic{equation}\alph{zahler}}
\newcounter{zahler}
\setcounter{zahler}{1}
\beq
 \label{MSUSY}
Z_{L = 1}
= \frac{1}{2} \mbox{ Str } \ln \frac{\Delta}{\Delta_0}
\scs
\eeq
with
\addtocounter{equation}{-1}
\addtocounter{zahler}{1}
\beq
\Delta  =
\left( \ba{cc} A & \sqrt{2/\mu}~\ol \Gamma \gamma_5 B \gamma_5 \\
\sqrt{2 \mu} ~ \Gamma & B \gamma_5 B \gamma_5 \ea \right) .\label{MSUSY1}
\eeq \setcounter{zahler}{0}
In many cases (as the Yukawa theory to be considered below), this
procedure brings the supermatrix differential operator
$\Delta$ into the form
\beq
\Delta = - D_\mu D_\mu + Y ~,\label{Dsquare}
\eeq
where $D_\mu = \del_\mu + X_\mu$ and $X_\mu$, $Y$ are matrices in
superspace. Generalizing the ordinary heat-kernel expansion to
superspace, we will extract the second Seeley--DeWitt coefficient of the
generic operator (\ref{Dsquare}) to obtain the one-loop divergences
including the terms bilinear in fermion fields.

In order to avoid the pitfalls with $\gamma_5$ in $d$ dimensions, the 
steps leading from Eq. (\ref{Zss}) to (\ref{Dsquare}) may be 
carried through by using an intermediate regularization that allows
the evaluation of determinants in four dimensions. An example of such
a regularization is provided by Eq. (\ref{ptf}), where the integration over
the parameter $\tau$ may be cut off at the lower end. In the following,
we assume that $X_\mu,Y$ are independent of $\gamma_5$, as is the
case  in the Yukawa theory, and we return to dimensional regularization 
for ease of comparison with the standard methods to evaluate the loop 
integrals. In cases where $\gamma_5$ pertains (as in pion--nucleon 
effective theories), one may stick to the regularization just mentioned.

In the proper-time formulation, the one-loop functional assumes the
form 
\beqa
Z_{L = 1} &=& - \dfrac{1}{2} \int_0^\infty \dfrac{d\tau}{\tau} \mbox{ Str } 
\left(e^{-\tau \Delta}-e^{-\tau \Delta_0}\right) \nn
&=&  - \dfrac{1}{2} \int_0^\infty \dfrac{d\tau}{\tau} \int d^dx 
\mbox{ str } 
\langle x|e^{-\tau \Delta}-e^{-\tau \Delta_0}|x \rangle ~. 
\label{ptf}
\eeqa
To extract
the divergences, we need the coefficient $a_2(x,x)$ in the heat-kernel
expansion
\beq
\langle x|e^{-\tau \Delta}|x \rangle  = (4\pi
\tau)^{-d/2}\sum_{n=0}^\infty 
\tau^n a_n(x,x)~.\label{hke}
\eeq
The divergent part $Z_{L = 1}^{\rm{div}}$ of the one-loop functional
is then given by
\beq
Z_{L = 1}^{\rm{div}} = \dfrac{1}{(4\pi)^2(d-4)}\int d^4x \mbox{ str }
\left[a_2(x,x)-a_2^0(x,x)\right]~.\label{Zdiv}
\eeq
The derivation of the diagonal super-Seeley--DeWitt coefficient
$a_2(x,x)$ is completely analogous to the ordinary case. Employing the
method of Ball \cite{Ball}, we write
\beqa
\langle x|e^{-\tau \Delta}|x \rangle &=& \int d^dk \langle x|e^{-\tau
\Delta}|k \rangle \langle k|x \rangle =
\int \dfrac{d^dk}{(2\pi)^d}e^{-ikx} e^{-\tau\Delta} e^{ikx} \nn
&=& \int \dfrac{d^dk}{(2\pi)^d}e^{-\tau (k^2-2ik_\mu D_\mu + \Delta)}
{\bf 1}~.
\eeqa
With the dimensionless integration variable $l=\sqrt{\tau}k$, we obtain 
from
(\ref{hke})
\beqa
a_2(x,x)&=&\int \dfrac{d^dl}{\pi^d}e^{-l^2}\left[\dfrac{1}{2}\Delta^2+
\dfrac{2}{3}l_\mu l_\nu (\Delta D_\mu D_\nu +D_\mu \Delta D_\nu +
D_\mu D_\nu \Delta )\right.\nn 
& & + \left.\dfrac{2}{3}l_\mu l_\nu l_\rho l_\sigma 
D_\mu D_\nu D_\rho D_\sigma\right] {\bf 1}
\eeqa
and therefore, after integration over $l$,
\beq
a_2(x,x)= \dfrac{1}{12}X_{\mu\nu}X_{\mu\nu}+\dfrac{1}{2}Y^2+
\left(-\dfrac{1}{6}D^2Y - \dfrac{1}{6}YD^2 + 
\dfrac{1}{3}D_\mu Y D_\mu\right){\bf 1}~,
\eeq
with
$$
X_{\mu\nu}=[D_\mu,D_\nu]=\del_\mu X_\nu - \del_\nu X_\mu +
[X_\mu,X_\nu]~.
$$
By partial integration, the last term does not contribute to the
divergence functional (\ref{Zdiv}) and we arrive at the final result
\beq
\int d^4x \mbox{ str }a_2(x,x)=\int d^4x \mbox{ str }\left(
\dfrac{1}{12}X_{\mu\nu}X_{\mu\nu}+\dfrac{1}{2}Y^2\right)~.
\eeq

\begin{figure}
\centerline{\epsfig{file=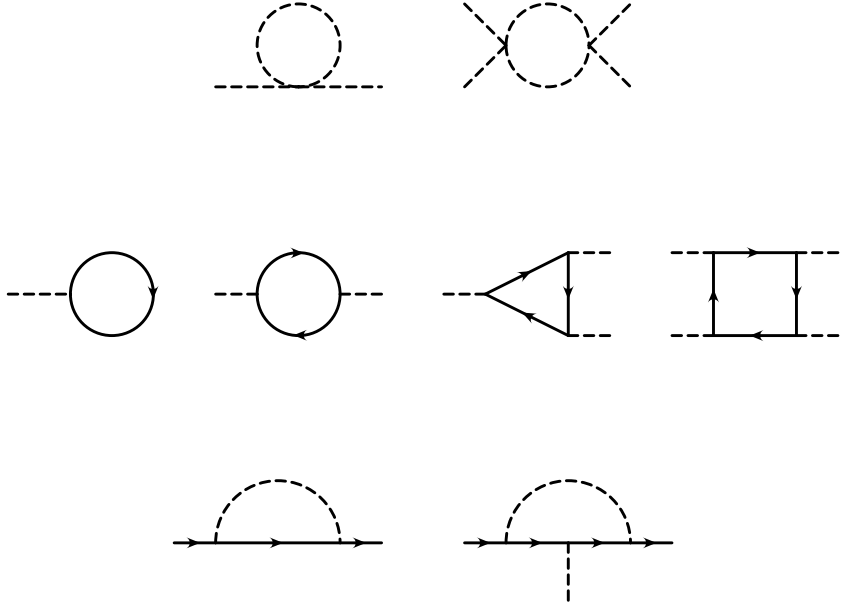,height=8cm}}
\caption{One-loop Feynman diagrams with divergent parts for the Yukawa
Lagrangian (\protect\ref{LYuk}). Bosons (fermions) are denoted by
dashed (full) lines.}
\label{fig:Yukawa}
\end{figure}

\paragraph{4.} We illustrate the formalism with a specific
case, the Yukawa theory of one scalar and one fermion field.
We consider the following Lagrangian in Euclidean space:
\beq
\cL = \frac{1}{2} \vp(- \partial^2 + M^2 )\vp 
+ \dfrac{\lambda}{4!}\vp^4 +
\ol \psi(\gamma_\mu \partial_\mu + m - g \vp)\psi - j \vp -
\ol \psi \rho - \ol \rho \psi~. \label{LYuk}
\eeq
The one-loop divergences of this theory are due to the Feynman diagrams in
Fig.~\ref{fig:Yukawa}: the first two diagrams are contained in the
bosonic one-loop functional, the next
four are included in the fermionic counterpart,
whereas the last two belong to the mixed functional, with both
bosons and fermions running in the loop.

The equations of motion are
\beqa
(- \partial^2 + M^2)\vp_{\rm cl} &=& j
- \dfrac{\lambda}{3!}\vp_{\rm cl}^3 + g \ol \psi_{\rm cl} \psi_{\rm
cl} \co\no \\
(\not\!\partial + m)\psi_{\rm cl} &=& \rho + g \vp_{\rm
cl}\psi_{\rm cl}\per\nonumber
 \eeqa
In terms of the fluctuation fields
defined in Eq. (\ref{ff}), the fluctuation Lagrangian is
\beq
\cL^{(2)} = \dfrac{1}{2}\xi(-\partial^2 + \wh M^2)\xi + 
\ol \eta(\not\!\partial + \wh m)\eta
- g\; \xi(\ol \psi_{\rm cl} \eta + \ol \eta \psi_{\rm cl})~,
\eeq
$$
\wh M^2 = M^2 + \dfrac{\lambda}{2}\vp_{\rm cl}^2 ~,
\qquad \wh m = m - g \vp_{\rm cl}~.
$$
Thus, the three entries of the supermatrix $K''$ in
Eq. (\ref{KPP}) are 
\beqa
A &=& - \partial^2 + \wh M^2\scs
B = \not\!\partial + \wh m \scs
\Gamma = - g \; \psi_{\rm cl} ~.\nonumber
\eeqa
The supermatrix $\Delta$ in Eq. (\ref{MSUSY1}) has
the structure (\ref{Dsquare}) with
\beq
X_\mu = \left( \ba{cc}
0 & - \dfrac{g}{\sqrt{2 \mu}} \; \ol \psi_{\rm cl} \gamma_\mu \\[7pt]
0 & 0 \ea \right), \quad
Y = \left( \ba{cc}
\wh M^2 & - \dfrac{g}{\sqrt{2 \mu}} \; \ol \psi_{\rm cl}
(\st{\gets}{\not\!\partial} + 2 \wh m) \\[7pt]
- \sqrt{2 \mu} \; g \psi_{\rm cl} & \not\!\partial \; \wh m + \wh m^2
\ea \right)~.
\eeq
In our simple example, without vector fields and without derivative
interactions, the curvature $X_{\mu\nu}$ 
does not contribute because $X_\mu X_\nu = 0$. The divergence Lagrangian
takes the form
\beqa
\cL^{\rm div} = \frac{1}{2(4\pi)^2(d-4)} \mbox{ str }
Y^2\scs\nonumber
 \eeqa
reducing the whole problem to a simple matrix multiplication. The final
result in Euclidean space is
\beqa
\cL^{\rm div} &=& \dfrac{1}{(4\pi)^2(d-4)} \left\{ 
\frac{1}{2}(\wh M^4 - M^4) - 2(\wh m^4 - m^4) 
- 2g^2 \partial_\mu \vp_{\rm cl} \partial_\mu \vp_{\rm cl}\right.\nn 
& & \left.\mbox{\hspace{2cm}} + g^2 \ol \psi_{\rm cl}
(- \not\!\partial + 2 \wh m) \psi_{\rm cl}
\right\}\per \label{LYdiv}
 \eeqa
An explicit calculation of the diagrams in Fig.~\ref{fig:Yukawa} 
reproduces exactly these divergences. The same result can be obtained 
with the background field method \cite{thooft}.

\paragraph{5.} We now present an alternative method for 
calculating the one-loop functional by applying the super-heat-kernel 
expansion directly to the operator $K'$ in (\ref{KP}) or to
$K''$ in (\ref{KPP}). For the Yukawa theory we are
considering here, the previous method is 
simple enough. However, for non-renormalizable Lagrangians
with derivative couplings occurring in chiral perturbation theory, the
expressions can become more compact in the following
approach. In addition, we wish to demonstrate with an explicit example
that the super-heat-kernel expansion is perfectly well defined also
in the linear version.

We confine ourselves to the mixed functional in this case,
i.e. to the part bilinear in the external fermion fields. Going through
essentially the same steps as in paragraph 3 and specializing
immediately to the Yukawa Lagrangian (\ref{LYuk}), we obtain
\beqa
Z_{L = 1} &=& - \dfrac{1}{2} \int_0^\infty \dfrac{d\tau}{\tau} \int d^dx 
\int \dfrac{d^dl}{(2\pi\sqrt{\tau})^d} \mbox{ str } 
(e^{V+W}{\bf 1}) + \ldots \label{Zlinear}
\eeqa
with
\beqa
V &=& - \mbox{diag}\left((l-i\sqrt{\tau}\del)^2 + \tau \wh M^2,\, 
\mu [i\sqrt{\tau}\!\fsl\,l +\tau(\fsl\del + \wh m)]\right)~,\nn
W &=& \sqrt{2 \mu}\tau g \left( \ba{cc}
0 & \ol \psi_{\rm cl}  \\
 \psi_{\rm cl} & 0
\ea \right) ~. \label{VW}
\eeqa
The appropriate decomposition of the exponential in (\ref{Zlinear})
can be performed by using  Feynman's ``disentangling'' theorem \cite{Feynman}:
\bea
\mbox{exp}(V+W) = \mbox{exp} ~ V ~ \mbox{P}\!_s \, \mbox{exp} 
\int_0^1 ds \, \wt{W} (s)                 
\eea
with
\bea
\wt{W} (s) := e^{-sV} W e^{sV} \no
\eea
and
\bea
\mbox{P}\!_s \, \mbox{exp} \int_0^1 ds \, \wt{W} (s) :=
\sum_{n=0}^{\infty} \int_0^1 ds_1 \int_0^{s_1} ds_2 \ldots \int_0^{s_{n-1}}
ds_n \, \wt{W} (s_1) \wt{W} (s_2) \ldots \wt{W} (s_n)~.
\eea
Since we are only interested here in terms with the structure
$\ol \psi_{\rm cl} \ldots \psi_{\rm cl}$, we pick out the part bilinear in
$W$:
\beq
\mbox{ str } (e^{V+W}{\bf 1}) = \int_0^1 ds \int_0^s ds'
\mbox{ str }\left[ e^{(1-s)V}W e^{(s-s')V}W e^{s'V} {\bf 1}\right] + \ldots 
\label{dtheorem}
\eeq
Manipulating this expression further, we will freely use partial
integration and shifts in $l$ keeping in mind that (\ref{dtheorem}) 
appears in the
one-loop functional (\ref{Zlinear}). Moreover, to avoid keeping track
of irrelevant terms, we use power counting to argue
[cf. Eq. (\ref{LYdiv})] that the divergences bilinear in the fermion
fields cannot contain any derivatives of $\wh M^2$ or $\wh m$ so that
we can treat  $\wh M^2$, $\wh m$ as constants. Of
course, $\wh M^2$ will not appear at all in the final result. With these
qualifications, we obtain for the terms of $\cO(\ol\psi \psi)$
\beqa
\mbox{ str } (e^{V+W} {\bf 1}) &=& 2 \mu \tau^2 g^2  \int_0^1 ds \int_0^s ds'
\left\{ e^{-(1-s+s')(l^2 + \tau \wh M^2)} \ol \psi_{\rm cl}
e^{-(s-s') \mu [i\sqrt{\tau}\!\fsl~l +\tau(\fsl\,\del + \wh m)]}\psi_{\rm
cl}\right.\nn
& & - \mbox{ tr } \left.\left[e^{-(1-s+s') \mu (i\sqrt{\tau}\!\fsl~l + \wh m)}
\psi_{\rm cl} e^{-(s-s')[(l-i\sqrt{\tau}\del)^2 + \tau \wh M^2]}
\ol \psi_{\rm cl}\right]\right\}\nn
&=& 2 \mu \tau^2 g^2  \int_0^1 ds \int_0^s ds' \, \ol \psi_{\rm cl}
\left\{e^{-(s-s') \mu (i\sqrt{\tau}\!\fsl~l +\tau \wh m)}
e^{-(1-s+s')[(l+i\sqrt{\tau}\del)^2 + \tau \wh M^2]}\right.\nn
& & \left. + e^{-(1-s+s') \mu (i\sqrt{\tau}\!\fsl~l +\tau \wh m)}
e^{-(s-s')[(l+i\sqrt{\tau}\del)^2 + \tau \wh M^2]}
\right\} \psi_{\rm cl} \nn
&=&  2 \mu \tau^2 g^2  \int_0^1 dz \, \ol \psi_{\rm cl}
e^{-z \mu (i\sqrt{\tau}\!\fsl~l +\tau \wh m)}
e^{-(1-z)[(l+i\sqrt{\tau}\del)^2 + \tau \wh M^2]}\psi_{\rm cl} ~.
\eeqa
After integration over $z$, the one-loop functional bilinear in the
external fermion fields is found to be
\beq
Z_{L = 1}|_{\ol\psi\psi} = \mu g^2\int d^dx \int_0^\infty \dfrac{d\tau}{\tau} 
\tau^{2-\frac{d}{2}} \int \dfrac{d^dl}{(2\pi)^d} \, \ol \psi_{\rm cl} 
\dfrac{e^{-[(l+i\sqrt{\tau}\del)^2 + \tau \wh M^2]}-
e^{- \mu (i\sqrt{\tau}\!\fsl~l +\tau \wh m)}}{(l+i\sqrt{\tau}\del)^2 + \tau
\wh M^2 - \mu (i\sqrt{\tau}\!\fsl\,l + \tau \wh m)} \psi_{\rm cl}~.
\label{Zpsi}
\eeq
Except for derivatives of $\wh M^2$ and $\wh m$, this is still the
complete one-loop functional bilinear in the fermion fields. 

To extract the divergent parts, we change integration variables and
decompose the functional (\ref{Zpsi}) in the following way:
\beqa
Z_{L = 1}|_{\ol\psi\psi} &=& g^2 \int d^dx  
\int \dfrac{d^dl}{(2\pi)^d} \, \ol \psi_{\rm cl} \left\{\mu
\int_0^\infty \dfrac{d\tau}{\tau} \,
\dfrac{\tau^{2-\frac{d}{2}}e^{-\tau \wh M^2}e^{-l^2}}{l^2 + 
\tau \wh M^2 - \mu [i\sqrt{\tau}(\!\fsl\,l - i\sqrt{\tau} \fsl\del)
 + \tau \wh m]}\right.\nn
& & \left. -  \int_0^\infty \dfrac{dt}{t} \,
\dfrac{t^{3-d}e^{- t \wh m}e^{-i\!\fsl~l}}
{(l + i t \del)^2 + t^2 \wh M^2 - \mu (i t\!\fsl\,l + t^2
\wh m)}\right\} \psi_{\rm cl}~. 
\eeqa
The first ($\mu$-dependent) term in this expression is cancelled by the
$\mu$-dependent parts of the second one.
The divergent parts are now easily isolated by expanding the second
denominator (setting $\mu = 0$) in $t$. With $l:=\sqrt{l^2}$, 
the divergent parts are given by
\beqa
Z_{L = 1}^{\rm div}|_{\ol\psi\psi} &=&  g^2\int d^dx \int_0^\infty 
\dfrac{dt}{t} e^{-t \wh m} t^{3-d}
\int \dfrac{d^dl}{(2\pi)^d} \, \ol \psi_{\rm cl}
\left( \dfrac{2 t \sin l}{d \, l^3}
\fsl\del - \dfrac{\cos l}{l^2}\right) \psi_{\rm cl}\nn
&\stackrel{d \ra 4}{\longrightarrow}& \dfrac{g^2}{(4\pi)^2(d-4)}
\int d^4x \, \ol \psi_{\rm cl}\left( - \fsl\del + 2 \wh m\right)
\psi_{\rm cl} ~.\label{lindiv}
\eeqa
For the final coefficients we have used the integrals
\beqa
\lim_{d\to 4} \int \dfrac{d^dl}{(2\pi)^d}\left(\dfrac{\cos l}{l^2},
\dfrac{\sin l}{l^3}\right) &=& \dfrac{1}{(4\pi)^2}(-2 , 2)~.\no
\eeqa
Comparing with the divergence Lagrangian (\ref{LYdiv}), we find
complete agreement with (\ref{lindiv}) in the fermionic part.

\paragraph{6.} In summary, we have shown that the one-loop
functional of a general quantum field theory, not necessarily of
the supersymmetric type, can be written in terms of the
Berezinian of a supermatrix
 operator. We have illustrated -- for the case of a Yukawa
theory -- that the divergent parts of the one-loop diagrams may
be worked out quite easily.
The method simplifies  in a  significant manner 
calculations in more realistic non-renormalizable theories \cite{NEUF98}.

\vspace{2cm}

\noindent


\end{document}